\documentclass[draftcls,onecolumn,a4paper]{IEEEtran}
\usepackage[cmex10]{amsmath}
\usepackage[psamsfonts]{amssymb}
\usepackage[dvipdf]{graphicx}
\usepackage{cite}
\usepackage{bm}

\newtheorem{theorem}{Theorem}
\newtheorem{lemma}{Lemma}

\newtheorem{definition}{Definition}
\newcommand{\eqtri}{\stackrel{\bigtriangleup}{=}}

\newcommand{\bX}{\bm{X}}
\newcommand{\bY}{\bm{Y}}
\newcommand{\bZ}{\bm{Z}}

\newcommand{\cS}{\mathcal{S}}

\newcommand{\cX}{\mathcal{X}}
\newcommand{\cY}{\mathcal{Y}}
\newcommand{\cZ}{\mathcal{Z}}
\newcommand{\tcZ}{\tilde{\cZ}}
\newcommand{\tz}{\tilde{z}}

\newcommand{\cJ}{\mathcal{J}}
\newcommand{\cA}{\mathcal{A}}
\newcommand{\cC}{\mathcal{C}}
\newcommand{\cD}{\mathcal{D}}

\newcommand{\cG}{\mathcal{G}}

\newcommand{\Int}{\mathbb{N}}
\newcommand{\Exp}{\mathbb{E}}

\newcommand{\abs}[1]{\left\lvert#1\right\rvert}
\newcommand{\funcabs}[1]{\left\lVert#1\right\rVert}
\newcommand{\bd}{\bar{d}}
\newcommand{\bk}{l}

\title{Universal Coding of Ergodic Sources for Multiple Decoders with Side Information}
\author{Shigeaki~Kuzuoka, Akisato Kimura, and Tomohiko Uyematsu
\thanks{S.~Kuzuoka is with the Department of Computer and Communication
Sciences, Wakayama University, 930 Sakaedani, Wakayama, 640-8510 Japan
(e-mail: kuzuoka@ieee.org)}%
\thanks{A.~Kimura is with NTT Communication Science Laboratories, NTT
Corporation, 3-1 Morinosato Wakamiya, Atsugi-shi, Kanagawa, 243-0198
Japan (e-mail: akisato@ieee.org)}%
\thanks{T.~Uyematsu is with the Department of Communications and
Integrated Systems, Tokyo Institute of Technology, 2-12-1 Ookayama,
Meguro-ku, Tokyo, 152-8550 Japan (e-mail: uyematsu@ieee.org)}%
\thanks{The work of S.~Kuzuoka was 
supported in part by Grant-in-Aid for Young Scientists (B) 22760278.
The work of A.~Kimura was supported in part by Grant-in-Aid for
Young Scientists (B) 20760255.}
}
\date{\today}

\begin{document}
\maketitle

\begin{abstract}
A multiterminal lossy coding problem, which includes 
various problems
such as the Wyner-Ziv problem and the complementary delivery problem
as special cases, is considered.  
It is shown that any point in the
achievable rate-distortion region can be attained even if the source
statistics are not known. 
\end{abstract}

\section{Introduction}
\begin{figure}[!p]
\centering
\includegraphics[width=2.2in]{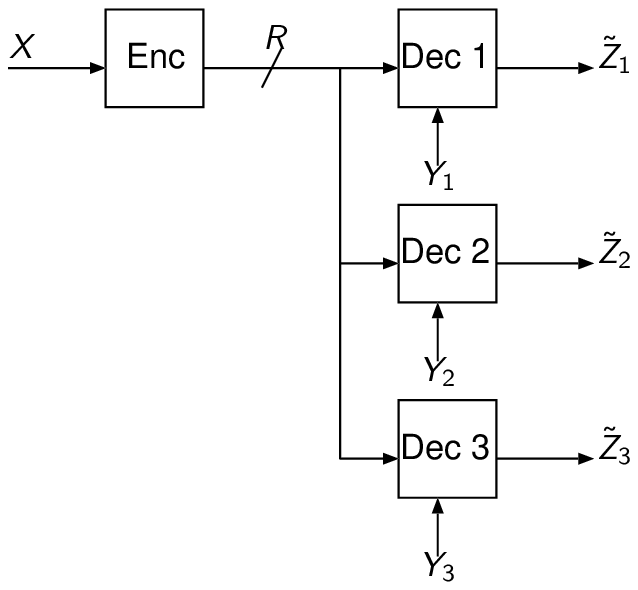}
\caption{Our coding system ($J=3$).} \label{fig1}
\end{figure}

Recently, the authors investigated the following
coding problem \cite{ISIT2010}.
Consider a coding system composed of one encoder and $J$ decoders.
The encoder observes the sequence
generated by a memoryless source with generic
variable $X$.
Then, the encoder broadcasts the codeword to the
decoders over the noiseless channel with capacity $R$.  
The purpose of the
$j$-th decoder is to estimate the value of the target source $Z_j$ as
accurately as possible by using the side information $Y_j$ and the codeword
sent by the encoder, where $\{Y_j\}_{j=1}^J$ and $\{Z_j\}_{j=1}^J$ may be correlated with $X$.
Accuracy of the estimation of the $j$-th decoder is evaluated by some
distortion measure $d_1^{(j)}$ and
it is required that the expected distortion is not greater than the
given value $\Delta_j$.
Fig.~\ref{fig1} depicts the coding system where $J=3$.

In \cite{ISIT2010}, we proposed a coding scheme which is
\emph{universal} in the sense that it attains the optimal
rate-distortion tradeoff even if the probability distribution $P_X$ of
the source $X$ is unknown, while the side informations $\{Y_j\}_{j=1}^J$
and the targets $\{Z_j\}_{j=1}^J$ are assumed to be generated from $X$
via a known memoryless channel.  In \cite{ISIT2010}, we considered only
stationary and memoryless sources.  In this paper, we extend the result
of \cite{ISIT2010} to the case where sources are stationary and ergodic
sources.

\begin{figure}[!p]
\centering
\includegraphics[width=2.2in]{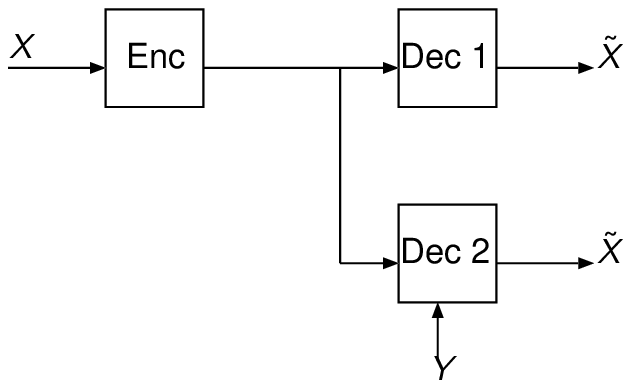}
\caption{Wyner-Ziv coding when side information may be absent.}\label{fig3}
\end{figure}

\begin{figure}[!p]
\centering
\includegraphics[width=2.2in]{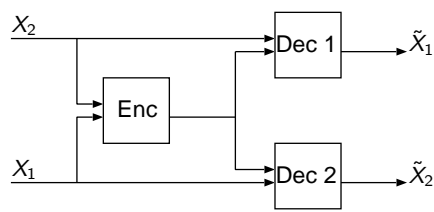}
\caption{Complementary delivery.}\label{fig2}
\end{figure}

\begin{figure}[!p]
\centering
\includegraphics[width=2.2in]{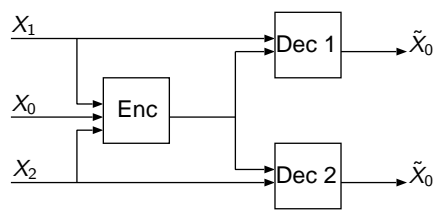}
\caption{Coding system considered in \cite{PerronDiggaviTelatar06}.} \label{fig4}
\end{figure}

As mentioned in \cite{ISIT2010}, our coding problem described above includes various problems as special
cases.  For example, the Wyner-Ziv problem, i.e.~the rate-distortion
problem with side information at the decoder \cite{WynerZiv76}, is a
special case of our problem, where $J=1$ and $Z_1=X$.  A variation of
the Wyner-Ziv problem, where the side information may fail to reach the
decoder \cite{HeegardBerger85,Kerpez87,Kaspi94}, is also included as a
special case, where $J=2$, $Y_1=\emptyset$, $Y_2=Y$ and $Z_1=Z_2=X$ (see
Fig.~\ref{fig3}).
Moreover, our coding system can be considered as a generalization of
the complementary delivery \cite{WynerWolfWillems02,KimuraUyematsuKuzuokaWatanabe09}.
In fact, a simple complementary delivery problem depicted in
Fig.~\ref{fig2} is the case where 
$J=2$, $X=\{X_1,X_2\}$, $Y_1=X_2$, $Y_2=X_1$, and $Z_j=X_j$ ($j=1,2$).
Further, our coding problem includes also the problem considered in
\cite{PerronDiggaviTelatar06} (depicted in Fig.~\ref{fig4}) as a special case, where $J=2$, $X=\{X_0,X_1,X_2\}$, $Y_j=X_j$
($j=1,2$), and $Z_1=Z_2=X_0$.

\section{Main Result}\label{sec:main}
At first, we introduce some notations.  We denote by $\Int$ the set of
positive integers $\{1,2,\dots\}$.  For a set $\cA$ and an integer
$n\in\Int$, $\cA^n$ denotes the $n$-th Cartesian product of $\cA$.  For
a finite set $\cA$, $\abs{\cA}$ denotes the cardinality of $\cA$.
Throughout this paper, we will take all $\log$ and $\exp$ to the base 2.


Let $\bX=X_1X_2\dots$ be a stationary and ergodic source with finite
alphabet $\cX$. For
each $n\in\Int$, $X^n$ denotes the first $n$ variables
$(X_1,X_2,\dots,X_n)$ of $\bX$ and the distribution of $X^n$ is denoted
by $P_{X^n}$.

Fix $J\in\Int$.  We consider random variables $Y_j$ (resp.~$Z_j$) taking
values in sets $\cY_j$ (resp.~$\cZ_j$) where $j$ ranges over the index
set $\cJ\eqtri\{1,2,\dots,J\}$.  We assume that, for each $j\in\cJ$,
$\cY_j$ and $\cZ_j$ are finite sets.  We write
\begin{align*}
 \cY_\cJ\eqtri\prod_{j\in\cJ}\cY_j,\quad
Y_\cJ\eqtri\{Y_j\}_{j\in\cJ}
\end{align*}
and
\begin{align*}
 \cZ_\cJ\eqtri\prod_{j\in\cJ}\cZ_j,\quad
 Z_\cJ\eqtri\{Z_j\}_{j\in\cJ}.
\end{align*}
 
Let $W\colon\cX\to\cY_\cJ\times\cZ_\cJ$ be a \emph{transition probability}.
In the followings, we assume that $W$ is
fixed and available as prior knowledge.
For each $n\in\Int$, let $W^n$ be the $n$-th extension of $W$, that
is,
\begin{equation*}
 W^n(y_\cJ^n,z_\cJ^n|x^n)\eqtri\prod_{i=1}^nW(y_{\cJ,i},z_{\cJ,i}|x_i)
\end{equation*}
for any sequences
\begin{align*}
x^n &\eqtri(x_1,x_2,\dots,x_n)\in\cX^n,\\
y_\cJ^n &\eqtri(y_{\cJ,1},y_{\cJ,2},\dots,y_{\cJ,n})\in\cY_\cJ^n,\text{ and}\\
z_\cJ^n &\eqtri(z_{\cJ,1},z_{\cJ,2},\dots,z_{\cJ,n})\in\cZ_\cJ^n. 
\end{align*}
Then, by a source $\bX$ and a transition probability $W$, sources
$\bY_\cJ\eqtri\{Y_{\cJ,i}\}_{i=1}^\infty$ and
$\bZ_\cJ\eqtri\{Z_{\cJ,i}\}_{i=1}^\infty$ are induced\footnote{Note that the transition probability $W$ is stationary and
\emph{memoryless}, while the source $\bX$ is stationary and ergodic.
Further, we assume that $W$ is \emph{known} both to the encoder and
decoders, while $\bX$ is unknown.
Universal Wyner-Ziv coding in a setting similar to ours is considered in \cite{JalaliVerduWeissman10}.
}.
In other words, 
$Y_\cJ^n\eqtri(Y_{\cJ,1},\dots,Y_{\cJ,n})$
(resp. $Z_\cJ^n\eqtri(Z_{\cJ,1},\dots,Z_{\cJ,n})$)
is a random variable on $\cY_\cJ^n$ (resp.~$\cZ_\cJ^n$) such that
\begin{align*}
\lefteqn{P_{X^nY_\cJ^nZ_\cJ^n}(x^n,y_\cJ^n,z_\cJ^n)}\\
&=P_{X^n}(x^n)W^n(y_\cJ^n,z_\cJ^n|x^n)\\
&=P_{X^n}(x^n)\prod_{i=1}^n W(y_{\cJ,i},z_{\cJ,i}|x_i)
\end{align*}
for any $x^n\in\cX^n$, $y_\cJ^n\in\cY_\cJ$, and $z_\cJ^n\in\cZ_\cJ^n$.
For each $j\in\cJ$, $\bY_j\eqtri\{Y_{j,i}\}_{i=1}^\infty$
(resp.~$\bZ_j\eqtri\{Z_{j,i}\}_{i=1}^\infty$) is called the $j$-th
component of $\bY_\cJ$ (resp.~$\bZ_\cJ$).
Note that the joint distribution $P_{X^nY_j^nZ_j^n}$ of $X^n$,
$Y_j^n\eqtri(Y_{j,1},\dots,Y_{j,n})$, and
$Z_j^n\eqtri(Z_{j,1},\dots,Z_{j,n})$ is given as a marginal distribution
of
$P_{X^nY_\cJ^nZ_\cJ^n}$, that is,
\begin{align*}
\lefteqn{
 P_{X^nY_j^nZ_j^n}(x^n,y_j^n,z_j^n)}\\
&\eqtri\sum_{\hat{j}\neq j}\sum_{y_{\hat{j}}^n,z_{\hat{j}}^n}
P_{X^nY_\cJ^nZ_\cJ^n}(x^n,y_\cJ^n,z_\cJ^n)
\end{align*}
for any $x^n\in\cX^n$, $y_j^n\in\cY_j^n$, and $z_j^n\in\cZ_j^n$, where
the summation is over all $(y_\cJ^n,z_\cJ^n)\in\cY_\cJ^n\times\cZ_\cJ^n$
such that the $j$-th component is $(y_j^n,z_j^n)$.

Further, for each $j\in\cJ$, let $\tcZ_j$ be a finite set.
Then, the formal definition of a code for our coding system is given as
follows.

\begin{definition}
An \emph{$n$-length block code}
\[
C_n=(\phi_n,\psi_n^{(1)},\dots,\psi_n^{(J)}) 
\]
is defined by mappings
\[
 \phi_n\colon\cX^n\to\{1,2,\dots,M_n\}
\]
and
\[
 \psi_n^{(j)}\colon\{1,2,\dots,M_n\}\times\cY_j^n\to\tcZ_j^n,\quad\forall j\in\cJ.
\]
$\phi_n$ is called the \emph{encoder} and $\psi_n^{(j)}$ is called the
\emph{$j$-th decoder}.
\end{definition}
\medskip

The performance of a code $C_n=(\phi_n,\psi_n^{(1)},\dots,\psi_n^{(J)})$
is evaluated by the coding rate and the distortion attained by $C_n$.
The \emph{coding rate} of $C_n$ is defined by
$(1/n)\log\funcabs{\phi_n}$, where $\funcabs{\phi_n}$ is the number
$M_n$ of the codewords of $C_n$.  For each $j\in\cJ$, let
\[
 d_1^{(j)}\colon\tcZ_j\times\cZ_j\to[0,d_{\max}^{(j)}]
\]
be a \emph{distortion measure}, where $d_{\max}^{(j)}<\infty$.  Then,
for each $n\in\Int$, the distortion between the output
$\tz_j^n=(\tz_{j,1},\dots,\tz_{j,n})\in\tcZ_j^n$ 
of the $j$-th decoder 
and the sequence
$z_j^n=(z_{j,1},\dots,z_{j,n})\in\cZ_j^n$ to be estimated is
evaluated by
\[
 d_n^{(j)}(\tz_j^n,z_j^n)\eqtri\frac{1}{n}\sum_{i=1}^nd_1^{(j)}(\tz_{j,i},z_{j,i}).
\]

\begin{definition}
 A pair $(R,\Delta_\cJ)$ of a rate $R$ and a $J$-tuple
 $\Delta_\cJ=(\Delta_1,\dots,\Delta_J)$ of distortions is said to be
 \emph{achievable} for a source $\bX$ if the
 following condition holds: For any $\epsilon>0$ and sufficiently large
 $n$ there exists a code
 $C_n=(\phi_n,\psi_n^{(1)},\dots,\psi_n^{(J)})$ satisfying
\begin{equation*}
 \frac{1}{n}\log\funcabs{\phi_n}\leq R+\epsilon
\end{equation*}
and, for any $j\in\cJ$,
\begin{align*}
\Exp_{X^nY_j^nZ_j^n}
\left[
  d_n^{(j)}\left(\psi_n^{(j)}(\phi_n(X^n),Y_j^n),Z_j^n\right)
\right]
\leq\Delta_j+\epsilon
\end{align*}
where $\Exp_{X^nY_j^nZ_j^n}$ denotes the expectation with respect to the
 distribution $P_{X^nY_j^nZ_j^n}$.
\end{definition}
\medskip

Now, we state our main result. The theorem clarifies that, whenever
$(R,\Delta_\cJ)$ is achievable, $(R,\Delta_\cJ)$ is also achievable
universally.

\begin{theorem}
\label{maintheorem}
For given $(R,\Delta_\cJ)$ and $\delta>0$, there exists a sequence $\{\bar{C}_n\}_{n=1}^\infty$ of codes
which is universally optimal in
 the following sense:
For any source $\bX$ for which 
$(R,\Delta_\cJ)$ is achievable 
there exists $n_0=n_0(\delta,\bX)$ such that, for any $n\geq n_0$,  $\bar{C}_n=(\bar\phi_n,\bar\psi_n^{(1)},\dots,\bar\psi_n^{(J)})$ 
 satisfies
\begin{equation*}
 \frac{1}{n}\log\funcabs{\bar\phi_n}\leq R+\delta
\end{equation*}
and 
\begin{equation*}
\Exp_{X^nY_j^nZ_j^n}
\left[
  d_n^{(j)}\left(\bar\psi_n^{(j)}(\bar\phi_n(X^n),Y_j^n),Z_j^n\right)
\right]\leq\Delta_j+\delta
\end{equation*}
for any $j\in\cJ$.
\end{theorem}
\medskip

The proof of the theorem will be given in the next section.

\section{Proof of Theorem \ref{maintheorem}}
Let $(R,\Delta_\cJ)$ and $\delta>0$ be given.
Fix $\epsilon>0$ satisfying
\[
 4J\epsilon+2\epsilon D_{\max}\leq\delta
\]
where
\begin{align*}
D_{\max}\eqtri\max_{j\in\cJ}d_{\max}^{(j)}.
\end{align*}
For each $n\in\Int$, let
$k_n\eqtri\log\log n$.

Let $\cC_n$ be the set of all $n$-length block codes
$C_n=(\phi_n,\psi_n^{(1)},\dots,\psi_n^{(J)})$ such that
$\funcabs{\phi_n}\leq 2^{n(R+\epsilon)}$.
Then, let $\cD_n$ be the set of $J$-tuple
$(\psi_n^{(1)},\dots,\psi_n^{(J)})$ of decoders such that 
$(\phi_n,\psi_n^{(1)},\dots,\psi_n^{(J)})\in\cC_n$ for some $\phi_n$.
Note that for $\bk\in\Int$,
\begin{equation}
 \abs{\cD_{\bk}}\leq \prod_{j\in\cJ}\left(\abs{\tcZ_j}^{\bk}\right)^{(2^{\bk(R+\epsilon)}\abs{\cY_j}^{\bk})}.
\label{eq:code_index_negligible}
\end{equation}

For each $j\in\cJ$, a sequence $x^n\in\cX^n$, and a code
$C_n=(\phi_n,\psi_n^{(1)},\dots,\psi_n^{(J)})$, let
\begin{align*}
\lefteqn{
 \bd_n^{(j)}(x^n,C_n)
}\\
&\eqtri
\sum_{y_j^n,z_j^n}P_{Y_j^nZ_j^n|X^n}(y_j^n,z_j^n|x^n)d_n^{(j)}\left(\psi_n^{(j)}(\phi_n(x^n),y_j^n),z_j^n\right).
\end{align*}
It should be noted that, by using $\bd_n^{(j)}(x^n,C_n)$, the average distortion attained by the code
$C_n$ can be written as
\begin{align}
\lefteqn{
 \Exp_{X^nY_j^nZ_j^n}
\left[
  d_n^{(j)}\left(\psi_n^{(j)}(\phi_n(X^n),Y_j^n),Z_j^n\right)
\right]}\nonumber\\
&=
\sum_{x^n}P_{X^n}(x^n)\sum_{y_j^n,z_j^n}\biggl\{P_{Y_j^nZ_j^n|X^n}(y_j^n,z_j^n|x^n)\nonumber\\
&\qquad\times
d_n^{(j)}\left(\psi_n^{(j)}(\phi_n(x^n),y_j^n),z_j^n\right)\biggr\}\nonumber\\
&=\sum_{x^n}P_{X^n}(x^n)\bd_n^{(j)}(x^n,C_n).\label{eq:property_of_bd}
\end{align}

For $\bk$ ($1\leq \bk\leq n$) and $s$ ($0\leq s<\bk$), let $q_{\bk;s}$ be the
\emph{non-overlapping empirical distribution} of $x^n$ defined as
\[
 q_{\bk;s}(a^\bk|x^n)\eqtri\frac{\abs{\{0\leq i< \lfloor
 (n-s)/\bk\rfloor:x_{i\bk+1+s}^{(i+1)\bk+s}=a^k\}}}{\lfloor
 (n-s)/\bk\rfloor},\quad a^\bk\in\cX^\bk.
\]

For each $n\in\Int$, let $\cG_n$ be the set of all sequences
$x^n\in\cX^n$ satisfying the following condition:
There are an integer $\bk$ ($1\leq\bk\leq k_n$) and a code
$C_\bk\in\cC_\bk$ such that for some integer $s$ ($0\leq s<\bk$),
\begin{equation}
 \sum_{a^\bk\in\cX^\bk}
 q_{\bk;s}(a^\bk|x^n)
\bd_\bk^{(j)}(a^\bk,C_\bk)\leq
 \Delta_j+4J\epsilon,\quad\forall j\in\cJ.
\label{eq:find_code} 
\end{equation}

Now, we describe the construction of the code
$\bar{C}_n=(\bar{\phi}_n,\bar{\psi}_n^{(1)},\dots,\bar{\psi}_n^{(J)})$.

\begin{itemize}
 \item \emph{Encoder $\bar\phi_n$:}
The encoder encodes a given  sequence $x^n\in\cX^n$ as follows.
\begin{enumerate}
 \item If $x^n\in\cG_n$, then choose integers $\bk,s$ and a code
       $C_\bk$ satisfying \eqref{eq:find_code}.
If $x^n\notin\cG_n$ then error is declared\footnote{In this case, the
       encoder may choose a codeword arbitrarily and send it to the decoders.
The choice of the codeword, which is sent when the error is declared,
       does not affect the analysis of the distortion.}.
 \item Send $\bk$ and $s$ by using $2\log k_n$ bits.
 \item Send the index of decoders
$(\psi_{\bk}^{(1)},\dots,\psi_{\bk}^{(J)})\in\cD_{\bk}$
by using
$\log\abs{\cD_{\bk}}$ bits.
 \item Send the codewords $\phi_{\bk}(x_{i\bk+1+s}^{(i+1)\bk+s})$ of blocks
       $x_{i\bk+1+s}^{(i+1)\bk+s}$ ($0\leq i< \lfloor (n-s)/\bk\rfloor$)
encoded by $\phi_{\bk}$.
\end{enumerate}

 \item \emph{Decoder $\bar\psi_n^{(j)}$:}
The $j$-the decoder decodes
the received codeword as follows.
\begin{enumerate}
 \item Decode the first $2\log k_n$ bits of the received codeword and
       obtain $\bk$ and $s$.
 \item Decode the first $\log\abs{\cD_{\bk}}$ bits of the remaining part
       of the received codeword and
       obtain the decoders
$(\psi_{\bk}^{(1)},\dots,\psi_{\bk}^{(J)})\in\cD_{\bk}$
chosen by the encoder.
 \item Decode the remaining part of the received codeword by using 
$\psi_{\bk}^{(j)}$ and the side information $y_j^n$.
Then, the blocks
$\tz_{i\bk+1+s}^{(i+1)\bk+s}$ ($0\leq i< \lfloor (n-s)/\bk\rfloor$)
are obtained.
\end{enumerate}
The remaining part of the output $\tz^n$, i.e.~$\tz_1^{s}$ and $\tz_{\lfloor (n-s)/\bk\rfloor\bk+1+s}^n$, is defined
arbitrarily. Note that the total length of 
$\tz_1^{s}$ and $\tz_{\lfloor (n-s)/\bk\rfloor\bk+1+s}^n$ is at most $2\bk$.
\end{itemize}

\subsection{Optimality of the Code}

By the fact that 
$1\leq \bk\leq k_n$ and
$\cD_{\bk}$ satisfies \eqref{eq:code_index_negligible},
it is easy to see that the coding rate
$(1/n)\log\funcabs{\bar\phi_n}$ of $\bar{C}_n$ satisfies
\[
 \frac{1}{n}\log\funcabs{\bar\phi_n}\leq R+\delta
\]
for sufficiently large $n$.
Hence, to show the optimality of the code $\bar{C}_n$, it is sufficient
to bound the distortion attained by $\bar{C}_n$.

At first, suppose $x^n\in\cG_n$.
By \eqref{eq:find_code} and 
the additivity of the distortion measures, 
the code
$\bar{C}_n$ satisfies that
\begin{align*}
\bd_n^{(j)}(x^n,\bar{C}_n)
&\leq
\frac{1}{\lfloor (n-s)/\bk\rfloor}\sum_{i=0}^{\lfloor
 (n-s)/\bk\rfloor}\bd_{\bk}^{(j)}\left(x_{i\bk+1+s}^{(i+1)\bk+s},C_{\bk}\right)+\frac{2\bk d_{\max}^{(j)}}{n}\\
&=\sum_{a^\bk\in\cX^\bk}q_{\bk;s}(a^\bk|x^n)\bd_{\bk}^{(j)}\left(a^\bk,C_{\bk}\right)
+\frac{2\bk d_{\max}^{(j)}}{n}\\
&\leq \Delta_j+4J\epsilon+\frac{2k_n d_{\max}^{(j)}}{n}.
\end{align*}
for any $j\in\cJ$.

On the other hand, if $x^n\notin\cG_n$ then the error is declared (and the
codeword is chosen arbitrarily). In this case, the distortion occurred at the
$j$-th decoder is upper bounded by $d_{\max}^{(j)}$.

Hence, for any $j\in\cJ$, we have
\begin{align}
\lefteqn{\sum_{x^n}P_{X^n}(x^n)\bd_n^{(j)}(x^n,\bar{C}_n)}\nonumber\\
&=\sum_{x^n\in\cG_n}P_{X^n}(x^n)\bd_n^{(j)}(x^n,\bar{C}_n)\nonumber\\
&\qquad+\sum_{x^n\notin\cG_n}P_{X^n}(x^n)\bd_n^{(j)}(x^n,\bar{C}_n)\nonumber\\
&\leq\sum_{x^n\in\cG_n}P_{X^n}(x^n)\left(\Delta_j+4J\epsilon+\frac{2k_n d_{\max}^{(j)}}{n}\right)\nonumber\\
&\qquad+\sum_{x^n\notin\cG_n}P_{X^n}(x^n)d_{\max}^{(j)}\nonumber\\
&\leq\Delta_j+4J\epsilon+\frac{2k_n d_{\max}^{(j)}}{n}+P_{X^n}\left(\cG_n^{\complement}\right) d_{\max}^{(j)}\nonumber\\
&\leq\Delta_j+4J\epsilon+\frac{2k_n D_{\max}}{n}+P_{X^n}\left(\cG_n^{\complement}\right) D_{\max}\label{eq:bound1}
\end{align}
where $\cG_n^{\complement}$ denotes the complement of $\cG_n$.

Further, as shown in Lemma \ref{lemma:keylemma} in the appendix,
if $(R,\Delta_\cJ)$ is achievable for $\bX$ then
\begin{equation}
 P_{X^n}\left(\cG_n^\complement\right)\leq\epsilon
\label{eq:target}
\end{equation}
holds for sufficiently large $n$. 

Hence, for sufficiently large $n$,
\begin{equation}
 \sum_{x^n}P_{X^n}(x^n)\bd_n^{(j)}(x^n,\bar{C}_n)
\leq \Delta_j+4J\epsilon+2\epsilon D_{\max}\label{eq:bound2}
\end{equation}

By \eqref{eq:property_of_bd} and \eqref{eq:bound2}, the average distortion attained
by
$\bar{C}_n=(\bar{\phi}_n,\bar{\psi}_n^{(1)},\dots,\bar{\psi}_n^{(J)})$
is bounded as
\begin{equation*}
  \Exp_{X^nY_j^nZ_j^n}
\left[
  d_n^{(j)}\left(\bar\psi_n^{(j)}(\bar\phi_n(X^n),Y_j^n),Z_j^n\right)
\right]\leq \Delta_j+\delta
\end{equation*}
for any $j\in\cJ$.
This completes the proof of Theorem \ref{maintheorem}.

\appendix
\begin{lemma}
\label{lemma:keylemma}
Let $\bX$ be a stationary and ergodic source for which $(R,\Delta_\cJ)$ is achievable.
Then, for sufficiently large $n$, 
\[
 P_{X^n}\left(\cG_n^\complement\right)\leq\epsilon
\]
holds.
\end{lemma}

\begin{IEEEproof}
Since $(R,\Delta_\cJ)$ is achievable for $\bX$, 
there are an integer $\bk$ and a code $C_\bk$ such that
\begin{equation}
 (1/\bk)\log\funcabs{\phi_{\bk}}\leq R+\epsilon
\label{eq:2}
\end{equation}
and
\begin{equation}
  \Exp_{X^{\bk}Y_j^{\bk}Z_j^{\bk}}
\left[
  d_{\bk}^{(j)}\left(\psi_{\bk}^{(j)}(\phi_{\bk}(X^{\bk}),Y_j^{\bk}),Z_j^{\bk}\right)
\right]
\leq\Delta_j+\epsilon,\quad\forall j\in\cJ.
\label{eq:1} 
\end{equation}

For each $j\in\cJ$, let $f^{(j)}$ be a function on $\cX^\bk$ such that
\[
 f^{(j)}(a^\bk)\eqtri\bd_{\bk}^{(j)}(a^\bk,C_\bk)-\Delta_j-\epsilon,\quad a^\bk\in\cX^\bk.
\]
Then, \eqref{eq:1} implies that
\begin{equation*}
 \Exp_{X^{\bk}}
\left[
  f^{(j)}(X^\bk)
\right]
\leq 0.
\end{equation*}
Hence, the \emph{ergodic theorem} guarantees the following fact:
There exists $n_1=n_1(j,\epsilon,\bk,\bX)$ such that for any $n\geq n_1$
 there exists a set $\cA_n\subseteq\cX^n$ satisfying
(i) $\Pr\{X^n\in\cA_n\}\geq 1-\epsilon$ and (ii) for any $x^n\in\cA_n$,
\begin{equation}
 \sum_{a^\bk\in\cX^\bk}p_k(a^\bk|x^n)f^{(j)}(a^\bk)\leq \epsilon
\label{eq:ergodic_theorem}
\end{equation}
where $p_\bk$ is the \emph{overlapping empirical distribution} of $x^n$
 defined as
\[
 p_\bk(a^\bk|x^n)\eqtri\frac{\abs{\{1\leq i\leq
 n-\bk+1:x_i^{i+\bk-1}=a^\bk\}}}{n-k+1},\quad a^\bk\in\cX^\bk.
\]

Note that $p_\bk$ and $q_{\bk;s}$ satisfy that
\[
 (n-\bk+1)p_\bk(a^\bk|x^n)=\sum_{s=0}^{\bk-1}\lfloor(n-s)/\bk\rfloor
 q_{\bk;s}(a^\bk|x^n),\quad a^\bk\in\cX^\bk.
\]
and thus
\begin{equation}
 \frac{(n-\bk+1)}{(n-\bk)}p_\bk(a^\bk|x^n)\geq\frac{1}{\bk}\sum_{s=0}^{\bk-1}
 q_{\bk;s}(a^\bk|x^n),\quad a^\bk\in\cX^\bk.
\label{eq:4}
\end{equation}

By \eqref{eq:ergodic_theorem} and \eqref{eq:4}, for $n\geq n_1$ and $x^n\in\cA_n$,
\begin{align*}
 \frac{1}{\bk}\sum_{s=0}^{\bk-1}
\sum_{a^\bk\in\cX^\bk}
 q_{\bk;s}(a^\bk|x^n)f^{(j)}(a^\bk)\leq 
\frac{(n-\bk+1)}{(n-\bk)}\epsilon.
\end{align*}

Now, let $\cS(j,n,x^n)$ be the set of all $s$ such that
\begin{equation}
 \sum_{a^\bk\in\cX^\bk}
 q_{\bk;s}(a^\bk|x^n)f^{(j)}(a^\bk)>
2J\frac{(n-\bk+1)}{(n-\bk)}\epsilon. 
\label{eq:3}
\end{equation}
Then, by the Markov lemma,
\[
 \abs{\cS(j,n,x^n)}\leq \frac{\bk}{2J}
\]
for $n\geq n_1$ and $x^n\in\cA_n$.
Further, let $\cS(n,x^n)$ be the set of all $s$ such that
\eqref{eq:3} holds for at least one $j\in\cJ$.
Then, for $n\geq n_2\eqtri\max_j n_1(j,\epsilon,\bk,\bX)$ and $x^n\in\cA_n$, we have
\[
 \abs{\cS(n,x^n)}\leq \frac{\bk}{2}.
\]
Thus, for $n\geq n_2$ and $x^n\in\cA_n$, there exists at least one $s$ such that 
\[
  \sum_{a^\bk\in\cX^\bk}
 q_{\bk;s}(a^\bk|x^n)f^{(j)}(a^\bk)\leq
2J\frac{(n-\bk+1)}{(n-\bk)}\epsilon,\quad\forall j\in\cJ.
\]

On the other hand, we can choose $n_3$ such that for any $n\geq n_3$,
$(n-\bk+1)/(n-\bk)\leq 3/2$.
Then, for any $n\geq n_4\eqtri\max\{n_2,n_3\}$ and $x^n\in\cA_n$, we have
\begin{align}
  \sum_{a^\bk\in\cX^\bk}
 q_{\bk;s}(a^\bk|x^n)\bd_{\bk}^{(j)}(a^\bk,C_\bk)
&\leq\Delta_j+\epsilon+2J\frac{(n-\bk+1)}{(n-\bk)}\epsilon\nonumber\\
&\leq\Delta_j+\epsilon+3J\epsilon\nonumber\\
&\leq\Delta_j+4J\epsilon.\label{eq:5}
\end{align}
In other words, 
if $n$ is so large that $n\geq n_4$ and $\bk\leq k_n$
then
for any $x^n\in\cA_n$ we can choose $\bk$, $C_\bk$, and $s$ satisfying 
\eqref{eq:2} and \eqref{eq:5}.
This means that $\cA_n\subseteq\cG_n$.
Hence, we have
\[
 P_{X^n}(\cG_n)\geq P_{X^n}(\cA_n)\geq 1-\epsilon.
\]
This completes the proof of the lemma.
\end{IEEEproof}



\end{document}